\documentclass[11pt]{article}
\usepackage{amssymb, amsmath, amsfonts, amsthm}
\usepackage{pstricks,pst-node,pst-plot,color}
\usepackage{fullpage}
\usepackage[section]{algorithm}
\usepackage[noend]{algpseudocode}

\newtheorem{theorem}{Theorem}[section]
\newtheorem{definition}[theorem]{Definition}
\newtheorem{claim}[theorem]{Claim}
\newtheorem{lemma}[theorem]{Lemma}
\newcommand{\comment}[1]{}

\def\NN{\ensuremath{\mathop{\mathtt{NN}}}}
\def\DFS{\ensuremath{\mathop{\mathtt{DFS}}}}
\def\fignspace{}

\title{Nearest Neighbor Network Traversal}
\author{David Pritchard}

\begin{document}
\maketitle \begin{abstract} A mobile agent in a network wants to
visit every node of an $n$-node network, using a small number of
steps. We investigate the performance of the following ``nearest
neighbor'' heuristic: always go to the nearest unvisited node. If
the network graph never changes, then from (Rosenkrantz, Stearns and
Lewis, 1977) and (Hurkens and Woeginger, 2004) it follows that
$\Theta(n \log n)$ steps are necessary and sufficient in the worst
case. We give a simpler proof of the upper bound and an example that
improves the best known lower bound.

We investigate how the performance of this heuristic changes when it
is distributively implemented in a network. Even if network edges
are allow to fail over time, we show that the nearest neighbor
strategy never runs for more than $O(n^2)$ iterations. We also show
that any strategy can be forced to take at least $n(n-1)/2$ steps
before all nodes are visited, if the edges of the network are
deleted in an adversarial way. \end{abstract}


\section{Introduction}
In this paper we consider a problem about a computer network. We use
a graph as our model: each computer is a vertex, and we join two
vertices $u, v$ by an edge whenever there is a direct communication
link joining $u$ and $v$. An \emph{agent} in the network is an
entity that inhabits one node at a time, and is allowed to move in
\emph{steps}, where a step consists of the agent leaving its current
position $u$ and entering a neighbor $v$ of $u.$

The main goal of this paper is to discuss the task of \emph{agent
node traversal}, which is to make the agent visit all of the nodes
of the network at least once. One practical application of traversal
is that the agent can collect information from every node (like a
census), without the need for any global network coordination. One
might also use traversal as a way of exploring an initially unknown
network.

One well-known technique for performing a traversal is to use
\emph{depth first search} (\DFS), but we claim that it is not
practical in all real-world settings. In a depth first search,
whenever the agent is adjacent to an unvisited node, it moves to
that node; whenever all adjacent nodes are visited, the agent
backtracks its path by one step. It is not hard to see that in a
network of $n$ nodes, after $2(n-1)$ steps, the network will be
traversed and the agent will have returned to its initial position.
Here's the problem: what if some edges of the network die? The agent
may try to backtrack along an edge that no longer exists. Although
we could restart \DFS\ every time this happens, this solution seems
somewhat inelegant and inefficient.

One simple alternative mechanism for graph traversal is the
following: the agent always travels to the closest unvisited node.
This heuristic, which we call the \emph{nearest neighbor} (\NN)
strategy, is the subject of our paper. It has been studied before
under the guise of an approximation algorithm for the traveling
salesman problem (that setting differs from ours only in that, at
the end, the agent must return to its initial position). For each $n
\geq 1,$ let the \emph{approximation ratio} of \NN\ be the least
upper bound on the ratio $COST(g)/COST(OPT)$ where $g$ is an \NN\
traversal on an $n$-node graph, $COST$ measures the number of steps,
and $OPT$ is the cheapest traversal of the graph. Abusing this
definition slightly, for a fixed graph $G$ and \NN\ traversal $g,$
we sometimes call $COST(g)/COST(OPT)$ an approximation ratio of \NN\
\emph{on $G.$}

The authors of \cite{approxtsp} proved that the approximation ratio
of \NN\ is at most $(\frac{1}{2}+o(1))\log_2 n,$ and gave an
infinite family of edge-weighted graphs where \NN\ has an
approximation ratio of at least $(\frac{1}{3}+o(1))\log_2 n.$ A
simpler (and non-weighted) family was later found by \cite{hurkins}
giving an approximation ratio of at least $(\frac{1}{4}+o(1))\log_2
n.$

\subsection{Our Contribution}
The authors of \cite{ccps} point out that the upper bound proof in
\cite{approxtsp} is ``technical.'' In Section \ref{sec1} we give a
simple and new proof, but where the approximation ratio
$\frac{1}{2}\log_2 n$ is replaced by $\ln n,$ which is slightly
worse. Our approach is to bound the number of long steps in a
traversal generated by \NN. At the conclusion of the paper we
mention another application of this technique. \emph{Remark: this
paper was submitted to a conference and a referee pointed out that
this proof technique can also be found in an analysis
\cite{alon92line} by Alon and Azar of the Imase-Waxman online
minimum Steiner tree heuristic.}

We also improve the best known lower bound and show a family of
(non-weighted) graphs upon which \NN\ has an approximation ratio of
at least $(\frac{1}{2}+o(1))\log_2 n.$ This appears in Section
\ref{sec2}. With slightly more work, and using the upper bound from
\cite{approxtsp}, this establishes that the approximation ratio of
\NN\ is $(\frac{1}{2}+o(1))\log_2 n.$

In Section \ref{sec-appl}, we analyze a simple distributed
implementation of the \NN\ heuristic. If there are no faults, then
the algorithm always visits all nodes within $O(n \log n)$ time.
However, as our introduction suggests, we are interested in what
happens when faults are allowed. To our knowledge, ours is the first
such analysis. We allow edges to be destroyed over time, but no
edges are ever added to the graph or restored. We prove an $O(n^2)$
upper bound on the time before the distributed \NN\ algorithm
terminates (i.e., until every node remaining in the agent's
connected component is visited).

Finally, in Section \ref{sec:games}, we give a result which
indicates that \NN\ is in some sense optimal. We show that for every
strategy that an agent could use, there is an edge failure pattern
which forces the agent to take at least $\tbinom{n}{2}$ steps. Hence
\NN\ uses the least number of steps (in the worst case) of any
heuristic, up to a constant factor. This is not true of \DFS\ with
the ``restart when you cannot backtrack'' modification; see Appendix
\ref{app:dfs}.

\section{Static Graphs}
In this section, we assume that the graph $G = (V, E)$ does not
change over time. For any two nodes $u$ and $v,$ their
\emph{distance} $d(u, v)$ denotes the minimum number of edges in any
$u$-$v$ path. The nearest neighbor heuristic is shown below
(Algorithm \ref{alg:nn}). The algorithm takes a cost function $c$ as
input, so if we only want to count the number of steps taken by the
agent, then we would take $c=d.$ We consider only symmetric cost
functions in this paper, in other words, we assume $c(u, v)=c(v, u)$
for all nodes $u, v \in V.$
\begin{algorithm}
\begin{algorithmic}
\State $v_1$ is the initial location of the agent \For{$i$ from 2 to
$n$} \State let $v_i$ be any node in $V \backslash \{v_1, \dotsc,
v_{i-1}\}$ such that
 $c(v_{i-1}, v_i)$ is minimized
\EndFor
\end{algorithmic}
\caption{{\NN($G, c$): produces a nearest neighbor
traversal}}\label{alg:nn}
\end{algorithm}

A \emph{traversal} is any permutation of the vertex set $V.$ We call
any sequence $(v_1, \dotsc, v_n)$ that can be produced by the above
algorithm a \emph{nearest neighbor traversal.}

\subsection{Upper Bound}\label{sec1}
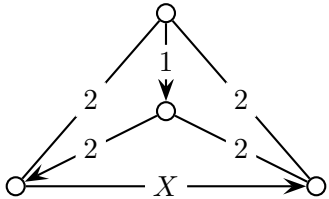
\begin{figure}
\begin{center} \leavevmode
\begin{pspicture}(-3.2, -1.2)(3.2,1.2)
 \psset{arrows=-}
 \cnodeput[](-2,-1){A}{$$}
 \cnodeput[](2,-1){B}{$$}
 \cnodeput[](0,0){C}{$$}
 \cnodeput[](0,1.3){D}{$$}
 \psset{labelsep=3pt}
 \ncline{A}{D}
 \lput*{U}{2}
 \ncline{B}{D}
 \lput*{U}{2}
 \ncline{B}{C}
 \lput*{N}{2}
 \psset{arrows=->,arrowsize=0.22cm}
 \ncline{D}{C}
 \lput*{U}{1}
 \ncline{C}{A}
 \lput*{N}{2}
 \ncline{A}{B}
 \lput*{N}{$X$}
\end{pspicture}
\caption{Take $X \to \infty$ in the diagram above. The cheapest
traversal has cost 5, but as indicated by the arrows, there is an
\NN\ traversal of cost $X+3.$ So without the triangle inequality,
the approximation ratio of \NN\ is unbounded. \fignspace}
\label{fig-trinec}
\end{center}
\end{figure}

A vertex sequence $ x= (x_1, \dotsc, x_k)$ has \emph{cost} $\lVert x
\rVert$ defined by
$$\lVert x \rVert := \sum_{i=1}^{k-1} c(x_{i}, x_{i+1}).$$
We say that a function $c$ satisfies the \emph{triangle inequality}
if $c(u, v) \leq c(u, w)+c(w, v)$ for all nodes $u, v, w.$ The
triangle inequality may be equivalent stated as: for a given $u$-$v$
path, replacing that path by the single edge $uv$ (sometimes called
\emph{short-cutting}) doesn't increase the cost. Note that $d$
satisfies the triangle inequality. The precise statement of the main
theorem of this section is as follows.
\begin{theorem}\label{thm1}
Suppose $c$ is a non-negative integer-valued function that is
symmetric and satisfies the triangle inequality. Then the
approximation ratio of $\NN(G, c)$ is at most $(1+\ln (n-1)).$
\end{theorem}
Now when $c=d$ (i.e., the cost is the number of agent steps) then a
\DFS\ traversal has cost at most $2(n-1),$ and hence in this case
Theorem \ref{thm1} implies that \NN\ always returns a traversal of
cost at most $O(n \log n).$ We remark that the triangle inequality
is necessary here to get a performance bound that depends only on
$n,$ as exemplified in Figure \ref{fig-trinec}.

\comment{By scaling, Theorem X also applies to rational-valued $d$:
multiply all edge weights by their common denominator $D,$ which
increases the cost of every traversal by a factor of $D,$ while
preserving the set of nearest neighbor traversals. Taking a suitable
limit, Theorem X holds for real-valued $d$ as well, but we do not
give all the details here.}

Hereafter we fix the number $n$ of vertices, a cost function $c$
that satisfies the triangle inequality, and a traversal $(g_1,
\dotsc, g_n)$ generated by \NN. Define $\lambda_j$ to be the number
of pairs of consecutive nodes with cost at least $j$ between them:
\begin{equation}\label{eqlam}
\lambda_j := |\{1 \leq i < n : c(g_i, g_{i+1}) \geq j \}|.
\end{equation}
By switching the order of summation, we see that the \NN\
traversal's cost is precisely the sum of all $\lambda_j:$
\begin{equation}\label{eqconj}
\lVert g \rVert = \sum_{i=1}^{n-1} c(g_i, g_{i+1}) =
\sum_{i=1}^{n-1} \sum_{j=1}^{c(g_i, g_{i+1})} 1 = \sum_{j \geq 1}
|\{1 \leq i < n : c(g_i, g_{i+1}) \geq j \}| = \sum_{j \geq 1}
\lambda_j.
\end{equation}

\begin{lemma} \label{lemma1}
Let $P = (P_1, \dotsc, P_k)$ be a partition of $V$ and suppose for
each $k,$ for any two nodes $u, v \in P_k,$ that $c(u, v)  \leq D.$
Then $\lambda_{D+1} \leq k-1.$
\end{lemma}
\begin{proof}
Suppose that $c(g_i, g_{i+1}) > D$ and let $g_i$ be contained in
class $P_j$ of the partition. We claim that $g_i$ was the last node
of $P_j$ visited by the agent. Otherwise, if $x \in P_j \backslash
\{g_1, \dotsc, g_i\},$ then after reaching $g_i$ the agent chose
$g_{i+1}$ such that $c(g_i,g_{i+1}) \geq D+1 > D \geq c(g_i, x),$
contradicting the fact that the agent makes greedy choices. Thus for
each of the $k$ parts $P_j$ there is at most one node $g_i$ in $P_j$
such that $c(g_i, g_{i+1}) > D.$ Furthermore, let $P_j$ be the part
containing $g_n,$ and we see $d(g_i, g_{i+1}) < D$ for each $g_i \in
P_j.$ The lemma then follows.
\end{proof}

Now, let $o = (o_1, \dotsc, o_n)$ be a traversal of optimal cost $C
:= \Vert o \Vert.$ By a short-cutting argument it is easy to see
that $c(o_i, o_j) \leq C$ for all vertices $o_i, o_j \in V$ and
hence $\lambda_j = 0$ when $j > C.$ Moreover, we obtain the
following bound on $\lambda_j$ for other values of $j.$

\begin{lemma}\label{lemma2}
For each positive integer $j$ we have $\lambda_j \leq C/j.$
\end{lemma}
\begin{proof}
By Lemma \ref{lemma1}, it suffices to exhibit a partition of $V$
into at most $C/j + 1$ parts, such that the pairwise costs within
each part are at most $j-1.$ We can do this by breaking $o$ into
paths of length about $j-1$ each. Define $a(1) = 1$ and iteratively
compute integers $a(i) \leq n$ for $i=2, 3, \dotsc $ such that
\begin{equation}\label{eq1}
\sum_{t = a(i)}^{a(i+1)-2} c(o_t, o_{t+1}) < j \leq \sum_{t =
a(i)}^{a(i+1)-1} c(o_t, o_{t+1}).
\end{equation}
This is continued as long as possible, that is, until some $k$
satisfies \begin{equation*}\label{eq2} \sum_{t=a(k)}^{n-1} c(o_t,
o_{t+1}) < j.
\end{equation*}
Note $a$ is a strictly increasing sequence, so $k$ is well-defined.
Let $a(k+1)=n+1$ and define $P_i := \{o_{a(i)}, \dotsc,
o_{a(i+1)-1}\}$ for $1 \leq i \leq k.$ Note that the $k$ sets $P_i$
partition $V.$ Furthermore, using the triangle inequality in a
short-cutting argument, it is easy to see that $c(u, v) < j$ for
each $u, v \in P_i.$

Intuitively, each $P_i$ accounts for a portion of $o$ of length $j,$
so we would expect $\lVert o \rVert / j = C/j$ parts plus a
remainder. Formally, using the definition of $C$ and Equation
\eqref{eq1}, we have
$$C \geq \sum_{t=1}^{a(k)-1} c(o_t, o_{t+1}) =
\sum_{i=1}^{k-1}\sum_{t=a(i)}^{a(i+1)-1} c(o_t, o_{t+1}) \geq
(k-1)j,$$ so $k,$ the number of parts, is at most $C/j+1,$ as
needed.
\end{proof}

Finally, we estimate the resulting bound on the length of $g,$ and
hence prove Theorem \ref{thm1}.
\begin{proof}[Proof of Theorem \ref{thm1}]
From Equation~\eqref{eqlam} we have $\lambda_i \leq n-1$ for all
$n.$ Further, recall that $\lambda_j = 0$ for $j > C.$ We may assume
without loss of generality that $n-1$ divides $C,$ as otherwise we
can increase $c$ uniformly by a factor of $n-1.$ We apply
Equation~\eqref{eqconj} and then Lemma \ref{lemma2}, obtaining
\begin{eqnarray*}
\lVert g \rVert = \sum_{j \geq 1} \lambda_j &\leq&
\sum_{j=1}^{C/(n-1)} (n-1)
+ \sum_{j=C/(n-1)+1}^{C} C/j \\
&\leq& (n-1) (C/(n-1)) +
C\sum_{j=C/(n-1) + 1}^{C} 1/j \\
&\leq& C + C \int_{C/(n-1)}^C dz/z \,\,\,\, = \,\,\,\, C(1+ \ln
(n-1)).
\end{eqnarray*}
Thus, as claimed, the cost of $g$ is at most $(1+\ln (n-1))$ times
the cost of $o.$
\end{proof}

Let $m$ (resp.\ $M$) denote the minimum (resp.\ maximum) value of
$c(u, v)$ over all pairs $\{u, v\} \subset V.$ We can tighten
Theorem \ref{thm1} in some cases and show that the approximation
guarantee of \NN\ depends logarithmically on the \emph{aspect ratio}
$\alpha := M/m:$
\begin{eqnarray*}
\lVert g \rVert = \sum_{j \geq 1} \lambda_j &\leq& \sum_{j=1}^{m}
(n-1)
+ \sum_{j=m+1}^{M} C/j \\
&\leq& m(n-1) + C \int_{m}^M dz/z\\
&\leq& C + C (\ln M - \ln m) \,\,\,\,=\,\,\,\, C(1 +  \ln \alpha).
\end{eqnarray*}
In comparison, Monnot \cite{monnot} showed that in the absence of
the triangle inequality, \NN\ has approximation ratio
$\frac{1+\alpha}{2}+o(1).$

\subsection{Lower Bound}\label{sec2} In this section we describe a
new family of graphs upon which \NN\ has an approximation ratio of
at least $(\frac{1}{2}+o(1))\log_2 n.$ We remark that the original
lower bound of \cite{approxtsp} could not be realized as the
distance function $d$ of any unweighted graph, but ours (like the
example from \cite{hurkins}) can be.

We call the family \emph{layered ring graphs} because of their
shape. The layered ring graphs are denoted $LR^k(2^m)$ where $k \geq
0$ is the number of layers and $m \geq 1$ is a size parameter. The
basic idea is that the agent in the nearest neighbor algorithm can
be forced to walk ``around" the ring $k$ times, once for each layer.

Each vertex in $LR^k(2^m)$ is assigned a \emph{position} $p(v) \in
\{0, 1, \dotsc, 2^m\}$ and we define two nodes $u, v$ to be adjacent
precisely when $p(v)-p(u) \in \{\pm 1, 0\} \pmod{2^m+1}.$ Every
layered ring graph includes the \emph{backbone} vertices $b_0, b_1,
\dotsc, b_{2^m}$ whose positions are $p(b_i) = i.$ It follows that
every layered ring graph is hamiltonian, since starting at $b_0$ we
can visit all vertices in position 0, then take an edge to $b_1$ and
subsequently visit all vertices in position 1, and so forth until we
return from the last vertex at position $2^m$ to $b_0.$

For notational convenience, we fix $m$ at this point and use only
$k$ as a parameter; so we omit $m$ and write $LR^k$ instead of
$LR^k(2^m).$ The first layered ring graph $LR^0$ consists of only
the backbone. Each ring graph $LR^k, k>0$ is constructed from the
previous one $LR^{k-1}$ by the addition of a \emph{layer}.

We'll show that on $LR^k,$ the agent can walk around the ring $k$
times, and hence the \NN\ heuristic can return a traversal of cost
about $k \cdot 2^m.$ As we will make precise in Lemma
\ref{lemma:lower}, when $k$ is roughly equal to $m/2,$ we'll have
$|V(LR^{m/2})| = 2^m(1+o(1));$ but since $LR^{m/2}$ is hamiltonian,
we get an approximation ratio of at least
$$\frac{COST(g)}{COST(OPT)} \approx \frac{k \cdot 2^m}{(1+o(1))2^m}
\approx k \approx \frac{1}{2}\log_2|V(LR^{m/2})|.$$

\begin{definition}
A \emph{layer} is a set $L$ such that $\{0, 2^m\} \subseteq L
\subseteq \{0, 1, \dotsc, 2^m\}$ with the following property: if
$a<b$ and $L \cap \{a, a+1, \dotsc, b\} = \{a, b\},$ then $b-a$ is a
power of $2.$
\end{definition}

We are about to define a sequence $L^1, L^2, \dotsc,$ of layers. We
say that $a, b$ are \emph{$L^i$-neighbors} if $L^i \cap \{a, \dotsc,
b\} = \{a, b\},$ and we denote this relation by $N^i(a, b)$ where
$a<b.$

\begin{definition} \label{deflay}
Define the first layer $L^1$ as follows:
$$L^1 := \{0, 1, 2, 4, \dotsc,
2^{m-1}, 2^m\}.$$ For $i \geq 1,$ define the $(i+1)$st layer
$L^{i+1}$ as follows:
$$L^{i+1} := \{0\} \cup \bigcup_{a, b : N^{i}(a, b)} \{a+2^t: 0 \leq t \leq \log_2(b-a)\}.$$
\end{definition}

Observe that $L^{i+1} \subseteq L^i$ for all $i \geq 1.$

\begin{definition} \label{lrdef} The layered ring graph $LR^0$ consists of the
backbone. For $i>0,$ the layered ring graph $LR^i$ consists of the
disjoint union of $LR^{i-1}$ together with one new vertex $\ell_t^i$
for each $t \in L^i.$ We define $p(\ell_t^i) = t.$
\end{definition}

\begin{figure}
\begin{center} \leavevmode
\begin{pspicture}(-3.8, -3.8)(3.8,3.8)
\input{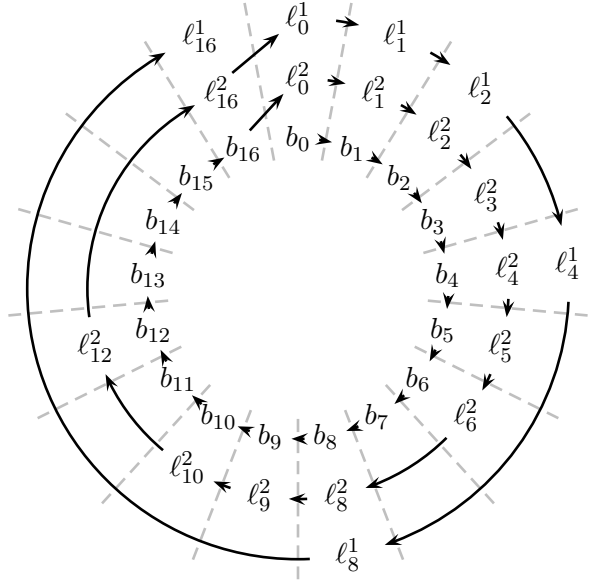}
\end{pspicture}
\caption{We illustrate the nodes of the layered ring graph
$LR^k(2^m)$ with $m=4, k=2.$ Two nodes are adjacent if their
positions (subscripts) differ by at most 1 modulo $(2^m+1).$ A \NN\
traversal of cost $(k+1)(2^m+1)-1$ beginning at $b_0$ is shown.
\fignspace }\label{figlr}
\end{center}\end{figure}

In Figure \ref{figlr}, we show an example of a layered ring graph.
For the rest of this section, $c$ is the distance function $d$ of
$LR^k.$

\begin{claim} It is possible for $\NN(LR^k, c)$ to return a traversal of
cost $(k+1)(2^m+1)-1.$ \label{mainclaim}
\end{claim}
\begin{proof}

We will show that the agent may visit the nodes in the following
order: backbone, layer $k,$ layer $k-1,$ and so on, visiting layer 1
last. Each layer, and also the backbone, is visited in increasing
order of position.

In the backbone, $c(b_i, b_{i+1})=1$ for $0 \leq i < 2^m,$ and so
these steps are valid for the \NN\ heuristic as all pairwise
distances are at least 1. Similarly, as $c(b_{2^m}, \ell^k_0)=1$ and
$c(\ell^i_{2^m}, \ell^{i-1}_0)=1$ for $k \geq i > 1,$ it remains
only to show that the intra-layer steps are valid. Precisely, for
each $k \geq i \geq 1$ and for each $a,b$ such that $N^{i}(a, b),$
say that a node $x \neq \ell^i_a$ is \emph{bad} if $c(\ell^i_a, x) <
c(\ell^i_a, \ell^i_b) = b-a$ and $x$ is not visited before
$\ell^i_a.$ Our goal is to show that no bad nodes exist.

Fix $a, b, i$ as above. First, consider the nodes of layer $i$ that
are not visited before $\ell^i_a.$ Since we visit layer $i$ in
increasing order of position, no node of the form $x = \ell^i_j, j
\geq b$ is bad unless $j > 2^m + 1 + a - (b - a).$ But from
Definition \ref{deflay} it follows easily that $b \leq 2a,$ and so
no bad nodes exist in layer $i.$

It remains to show that no bad nodes exist in layers $i-1, i-2,
\dotsc, 1.$ If $b=a+1$ then this is trivial. Otherwise, by the
definition of layers, it must be that $N^{i-1}(2a-b, b')$ holds
where $2a-b < a < b \leq b'.$ Thus there are no nodes in level $i-1$
with position strictly between $2a-b$ and $b,$ from which it follows
that level $i-1$ contains no bad nodes. Similarly since $L^j
\subseteq L^{i-1}$ for $j < i-1$ no bad nodes exist in any other
layer.
\end{proof}

In Appendix \ref{app:form}, we show a simple way to count the number
of nodes in $LR^k(2^m),$ obtaining:
\begin{equation}\label{form}
|V(LR^k(2^m))| = \comment{2^m + 1 + 3k +
\sum_{i=1}^{k}(2k+1-2i)\binom{m-1}{i} =} 2^m +
kO\biggl(\sum_{i=1}^k\binom{m-1}{i}\biggr).
\end{equation}

In what follows, we write $a \sim b$ to mean that $a = b(1+o(1)).$

\begin{lemma} \label{lemma:lower}
Fix $\delta > 0.$ For each $m,$ let $k = (m-1) / (2+\delta).$ Then
as $m \to \infty,$
$$|V(LR^k(2^m))| \sim 2^m.$$
\end{lemma}
\begin{proof}
First note that $|V(LR^k(2^m)| > 2^m.$ For an upper bound on
$|V(LR^k(2^m)|,$ Equation \eqref{form} gives
\begin{equation}\label{fooey}|V(LR^k(2^m)| = 2^m +
\frac{m-1}{2+\delta}O\Bigl(\sum_{i=1}^{(m-1)/(2+\delta)}\binom{m-1}{i}\Bigr).
\end{equation}
In turn, we can bound this expression by reinterpreting the sum
using a binomial random variable and applying a Chernoff bound
\cite[Thm. 4.2]{randalgs}. Namely,
$$\sum_{i=1}^{(m-1)/(2+\delta)}\binom{m-1}{i} = 2^m\Pr[Bin(m-1, 1/2)
\leq \frac{m-1}{2+\delta}] < 2^m
\exp\left(-\frac{(m-1)\delta^2}{4(2+\delta)^2}\right) =
o(2^mm^{-1}).$$ Then recalling Equation \eqref{fooey}, we see that
$|V(LR^k(2^m)| = 2^m + o(2^m)$ as claimed.
\end{proof}

\begin{theorem} \label{thm:lowerbound}
The approximation ratio $ar(n)$ of \NN\ satisfies
\begin{equation}
\limsup_{n \to \infty} ar(n) / \log_2(n) = 1/2. \label{eqaa}
\end{equation}
\end{theorem}
\begin{proof}
The upper bound in \cite{approxtsp} that we have mentioned implies
that the left-hand side of Equation~\eqref{eqaa} is at most 1/2. Now
pick any $\delta
> 0$ and consider the family of graphs in Lemma \ref{lemma:lower}.
The number $n$ of vertices satisfies $n \sim 2^m.$ By Claim
\ref{mainclaim} each graph in the family admits an \NN\ traversal of
cost $(2^m+1)(k+1)-1 \sim n \log_2(n)/(2+\delta).$ But each graph is
hamiltonian and so has optimal traversal cost $n-1.$ Hence $ar(n)
\gtrsim \log_2(n)/(2+\delta).$ As $m \to \infty$ so does $n \to
\infty,$ and by taking $m\to \infty$ and $\delta \to 0$ we obtain
Equation~\eqref{eqaa}.
\end{proof}

With a little more effort, we can replace the $\limsup$ in the above
equation by $\lim,$ or in other words we can establish that $ar(n)
\sim \frac{1}{2}\log_2 n.$ We defer the details to Appendix
\ref{app-fooey}.

\section{Network Implementation with Failures}\label{sec-appl}
As we stated in the introduction, one motivation for the nearest
neighbor algorithm is its potential usefulness in computer networks
with edge failures. We give a simple implementation below (Algorithm
\ref{alg2}). The variable $pos$ represents the position of the
agent. Each node $v$ keeps a flag $v.vis$ to indicate whether it has
been visited, and a number $v.dist$ which represents an estimate of
the distance from $v$ to the nearest unvisited node.
\begin{algorithm}
\begin{algorithmic}[1]
 \State set $v.dist := 0$ and $v.vis := false$ for each vertex
 \State let $pos$ be the agent's initial position, and set $pos.vis := true$
 \Loop
 \For{each node $v$ such that $v.vis = true,$ {\bf in parallel}}
 \State let $v.dist := 1 + \min\{u.dist \mid u=v \textrm{ or } uv \in E\}$ \label{line:dis1}
 \EndFor
 \If{$pos$ has a neighbor $u$ such that $u.dist < pos.dist$}
 \label{line:dis2}
 \State{
 set $pos := u$ and then set $pos.vis := true$}
 \EndIf
 \State some edges may be deleted
 \EndLoop
\end{algorithmic}
\caption{{Distributed Implementation of \NN\ Heuristic}}\label{alg2}
\end{algorithm}
Line \ref{line:dis1} determines the shortest paths to unvisited
nodes, and Line \ref{line:dis2} makes the agent travel along these
paths. For future reference, we need the following remarks:
\begin{enumerate}
\item[R1.] For each node $v,$ the value $v.dist$ is nondecreasing with
time.
\item[R2.] At all times, $v.dist$ is at most the actual distance to
the closest unvisited node.
\end{enumerate}
Remark R1 can be proved by induction on the number of iterations
elapsed, and remark R2 can be proved by induction on the distance to
the closest unvisited node.

 Say that a node is \emph{explored} the first time that the agent
visits it. If no failures occur, it is not too difficult to show
that the agent generates a greedy tour $g$ in the following way:
after exploring $g_i,$ it remains motionless for $d(g_i, g_{i+1})$
rounds, and in the following $d(g_i, g_{i+1})$ rounds it travels
directly to $g_{i+1}.$ Using Theorem \ref{thm1}, we find that
$\lVert g \rVert = O(n \log n),$ so all nodes are visited within
$O(n \log n)$ iterations. The purpose of this section is to show
that edge failures can dramatically increase the time complexity of
network traversal.

We considered a variant of the above implementation where each node
instantly knows the \emph{actual} distance to the nearest unvisited
node, but the results were essentially the same as what we present
here.

\subsection{Upper Bound}
If the graph becomes disconnected due to edge failures, then it may
not be possible for the agent to visit all of the nodes. Given this
fact, and furthermore that the agent may not initially know the
value of $n,$ how can we detect termination? We use the following
idea: the agent keeps a count $exp$ of how many nodes it has
explored so far, and once $pos.dist > exp,$ (using R2) there can be
no more reachable unvisited nodes. Using this as the definition of
termination, we now upper bound the algorithm's running time.
\begin{theorem}\label{thm:alg2fast}
Algorithm \ref{alg2} terminates in at most $O(n^2)$ iterations,
regardless of how the edge failures occur.
\end{theorem}
\begin{proof}
To simplify the arguments, suppose we do not permit any $dist$ label
to exceed $n+1$ (that is, once it hits this value, it does not
increase further). It is not hard to see that this does not affect
the observed behavior of the algorithm.

First, we claim there are at most $O(n^2)$ iterations in which the
agent moves. When the agent moves, the value $pos.dist$ decreases by
at least 1. However, the value $pos.dist$ can only increase $O(n^2)$
times, since $pos$ can be any of the $n$ nodes, and each node's
$dist$ label increases at most $n+1$ times (by R1).

Second, we can also show there are at most $O(n^2)$ iterations in
which the agent does not move.
\begin{claim} If the agent does not move in a given iteration, then
either the algorithm terminates in that iteration, or $v.dist$
increases for some node $v.$
\end{claim}
\begin{proof}
For the sake of contradiction, consider a non-final iteration in
which $v.dist$ does not increase for any node $v.$ By induction on
$t,$ we can show that every node $v$ at distance $t$ from the
nearest unvisited node has $v.dist = t,$ and if all nodes in the
connected component of $v$ are visited, then $v.dist = n+1.$ But
this is a contradiction, for it is easy to see that the agent would
have taken a step towards a nearest unvisited node.
\end{proof}
Since there are $n$ nodes and each nodes's $dist$ label can increase
at most $n+1$ times, we see that the agent remains still in $O(n^2)$
iterations.
\end{proof}
In Appendix \ref{app:dfs} we show, in contrast, that \DFS\ with
restarting (as described in the introduction) may take $\Omega(n^3)$
time.

\subsection{Lower Bound}\label{sec:games}
The upper bound of Theorem \ref{thm:alg2fast}, it turns out, has a
lower bound that matches it up to a constant factor. However, the
lower bound doesn't depend on any properties of the \NN\ heuristic.
Rather, we can show that \emph{any} heuristic for visiting all nodes
must take at least $\tbinom{n}{2}$ steps, if a suitable pattern of
edge deletions occurs.

We express this idea as a game: the objective of the agent is to be
in a connected component of $G$ where every node has been visited,
and an \emph{adversary} chooses the edges to delete, and wants to
foil the agent for as long as possible. An \emph{adaptive adversary}
--- one that can see the current state of the network in each
iteration before deciding what to delete --- is arguably the most
powerful adversary possible. We phrase our proof using an adaptive
adversary. Note however that for a \emph{deterministic} traversal
heuristic, a non-adaptive adversary is just as powerful as an
adaptive one, since the adversary can optimize its behavior ahead of
time by simulating the agent.

Here is what we, as the adversary, should do. The graph $G$ is
originally a complete graph on $n$ vertices. We wait until the agent
has visited $n-1$ nodes; let $v$ be the $(n-1)$st node visited, and
$x_1$ be the last remaining unvisited node. We then destroy the edge
$vx_1.$ As a result, the agent cannot visit $x_1$ in the next step.
Similarly, as soon as the agent moves to any other node $y$ such
that $y$ is adjacent to $x_1,$ we destroy the edge $yx_1.$ We
continue this until there are precisely 2 nodes $z_1$ and $z_1'$
adjacent to $x_1,$ and we wait for the agent to visit one or the
other (clearly the algorithm cannot terminate before then, since the
agent is connected to the unvisited node $x_1$). Without loss of
generality, assume the agent steps to $z_1$ before $z_1'.$ Then we
perform two edge deletions: we remove both $x_1z_1$ and $z_1z_1'.$
Define $x_2 := z_1'.$ Intuitively, we now want to keep the agent at
distance 2 or more from $x_1$ for as long as possible.

\begin{figure}
\begin{center} \leavevmode
\begin{pspicture}(-3.1, -1.8)(3.1,2.1)
\psset{unit=1.3cm}
 \psline*[linecolor=gray](-3.3, 0.0)(-2.7, 0.0)(-3, 0.35)
 \psline*[linecolor=gray](-3, -1.1)(-2.7, -0.6)(-3.3, -0.6)
 \pscircle*[linecolor=lightgray](-3, -0.3){0.4}
 \rput(-3, 2){$x_1$}
 \rput(-3, 1){$\hdots$}
 \rput(-3, 0.5){$x_i$}
 \rput(-3, 1.5){$x_2$}
 \psline(-3, 1.65)(-3, 1.85)
 \psline(-3, 1.15)(-3, 1.35)
 \psline(-3, 0.65)(-3, 0.85)
 \psline(-3.3, 0.0)(-3, 0.35)
 \comment{
 \psline(-3.1, 0.1)(-3, 0.35)
 \psline(-2.9, 0.1)(-3, 0.35)}
 \psline(-2.7, 0.0)(-3, 0.35)
 \psline(-3.3, -0.6)(-3, -1.1)
 \comment{\psline(-3.1, -0.7)(-3, -1.1)
 \psline(-2.9, -0.7)(-3, -1.1)}
 \psline(-2.7, -0.6)(-3, -1.1)
 \rput(-3, -1.1){\huge${{\star}}$}
 \psline[arrows=->,arrowsize=0.2,linearc=0.2](-3, -1.1)(-3.5,-0.7)(-3,-0.3)
 \psline[arrows=-*,arrowsize=0.2,linearc=0.2](-3, -1.1)(-3.5,-0.7)(-3,-0.3)
 \psline[linestyle=dashed,dash=2pt 2pt](-3,-0.3)(-3, 0.35)
 \rput(-2, 0.5){$\rightsquigarrow$}
 \psline*[linecolor=gray](-1.3, 0.0)(-0.7, 0.0)(-1, 0.35)
 \psline*[linecolor=gray](-1, -1.1)(-0.7, -0.6)(-1.3, -0.6)
 \psline*[linecolor=gray](-0.7, -1.1)(-0.7, -0.6)(-1.3, -0.6)
 \pscircle*[linecolor=lightgray](-1, -0.3){0.4}
 \rput(-1, 2){$x_1$}
 \rput(-1, 1){$\hdots$}
 \rput(-1, 0.5){$x_i$}
 \rput(-1, 1.5){$x_2$}
 \psline(-1, 1.65)(-1, 1.85)
 \psline(-1, 1.15)(-1, 1.35)
 \psline(-1, 0.65)(-1, 0.85)
 \psline(-1.3, 0.0)(-1, 0.35)
 \comment{
 \psline(-1.1, 0.1)(-1, 0.35)
 \psline(-0.9, 0.1)(-1, 0.35)}
 \psline(-0.7, 0.0)(-1, 0.35)
 \psline(-1.3, -0.6)(-1, -1.1)
 \comment{\psline(-1.1, -0.7)(-1, -1.1)
 \psline(-0.9, -0.7)(-1, -1.1)}
 \psline(-0.7, -0.6)(-1, -1.1)
 \psline(-1.3, -0.6)(-0.7, -1.1)
 \psline(-0.7, -0.6)(-0.7, -1.1)
 \rput(-0.7, -1.1){\huge$\star$}
 \psline[arrows=*-*](-1,-1.1)(-1,-1.1)
 \psline(-1,-1.1)(-0.7,-1.1)
 \psline[arrows=->,arrowsize=0.2,linearc=0.2](-0.7, -1.1)(-0.5,-0.7)(-1.1,-0.4)
 \psline[arrows=-*,arrowsize=0.2,linearc=0.2](-0.7, -1.1)(-0.5,-0.7)(-1.1,-0.4)
 \psline[linestyle=dashed,dash=2pt 2pt](-1, 0.35)(-1.1, -0.4)
 \rput(0, 0.5){$\hdots$}
 \psline*[linecolor=gray](0.65, -1.0)(0.7, -0.4)(1.35, -1.0)
 \psline*[linecolor=gray](0.65, -1.0)(1.3, -0.4)(1.35, -1.0)
 \psline(0.65, -1.0)(0.7, -0.4)(1.35, -1.0)
 \psline(0.65, -1.0)(1.3, -0.4)(1.35, -1.0)
 \pscircle*[linecolor=lightgray](1, -1){0.4}
 \rput(1, 2){$x_1$}
 \rput(1, 1){$\hdots$}
 \rput(1, 0.5){$x_i$}
 \rput(1, 1.5){$x_2$}
 \rput(0.7, -0.3){$z_i$}
 \rput(1.3, -0.3){$z'_i$}
 \psline[linestyle=dashed,dash=2pt 2pt](0.85, -0.3)(1.15, -0.3)
 \psline(1, 0.35)(1.3, -0.15)
 \psline[linestyle=dashed,dash=2pt 2pt](1, 0.35)(0.7, -0.15)
 \psline(1, 1.65)(1, 1.85)
 \psline(1, 1.15)(1, 1.35)
 \psline(1, 0.65)(1, 0.85)
 \rput(1, -1.1){\huge$\star$}
 \psline[arrows=->,arrowsize=0.2,linearc=0.2](1, -1.1)(0.3,-0.7)(0.55,-0.3)
 \rput(2, 0.5){$\rightsquigarrow$}
 \psline*[linecolor=gray](2.65, -1.0)(3.3, -0.4)(3.35, -1.0)
 \psline(2.65, -1.0)(3.3, -0.4)(3.35, -1.0)
 \psline*[linecolor=gray](2.7, -1.3)(3, -1.7)(3.3, -1.3)
 \psline(2.7, -1.3)(3, -1.7)(3.3, -1.3)
 \pscircle*[linecolor=lightgray](3, -1){0.4}
 \rput(3, 2){$x_1$}
 \rput(3, 1){$\hdots$}
 \rput(3, 0.5){$x_i$}
 \rput(3, 1.5){$x_2$}
 \rput(3.3, -0.3){$x_{i+1}$}
 \psline(3, 0.35)(3.3, -0.15)
 \psline[arrows=*->,linestyle=dashed,dash=2pt 2pt](3, -1)(3.3, -0.4)
 \psline(3, 1.65)(3, 1.85)
 \psline(3, 1.15)(3, 1.35)
 \psline(3, 0.65)(3, 0.85)
 \rput(3, -1.7){\huge$\star$}
 \psline[arrows=->,arrowsize=0.2,linearc=0.2](3, -1.7)(2.5,-1.4)(3,-1)
\end{pspicture}
\caption{A picture of phase $i.$ The agent (denoted by $\star$) is
trying to visit node $x_1,$ and the adversary wants to avoid this.
In each iteration, the agent moves to an adjacent node (denoted by
the arrow) and then the adversary deletes some edges (denoted by
dashed lines \fignspace). Shaded figures are cliques.}
\label{fig:phase}
\end{center}
\end{figure}
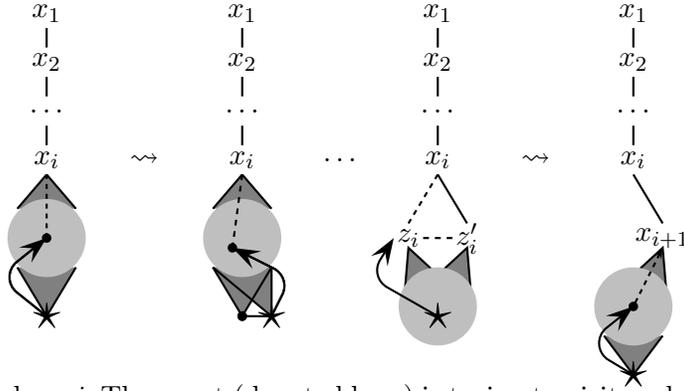

In general, the $i$th ``phase'' begins when $x_i$ is defined. Each
time the agent moves onto a node $y$ adjacent to $x_i,$ we delete
$yx_i.$ This continues until there are two nodes other than
$x_{i-1}$ adjacent to $x_i,$ which we call $z_i$ and $z'_i.$
W.o.l.o.g.\ let the agent reach $z_i$ first, and at that point, we
delete both $x_iz_i$ and $z_iz'_i.$ We also define $x_{i+1} := z'_i$
and the $(i+1)$st phase begins. We depict a generic phase in Figure
\ref{fig:phase}.

This can be continued until the end of the $(n-3)$rd phase, at which
point the nodes consist of $x_1, \dotsc, x_{n-3}, z'_{n-2}, y,
z_{n-2}$ with the agent at $z_{n-2}.$ The remaining graph is a path,
and the agent needs to take $(n-1)$ more steps to complete its
traversal. In the $i$th phase, the agent has to move onto $(n-i-2)$
distinct nodes. There are at least $(n-2)$ additional steps at the
beginning before we define $x_1.$ Hence the total number of steps is
at least
$$(n-2) + \bigl(\sum_{i=1}^{n-3} n-i-2 \bigr) + (n-1) =
\tbinom{n}{2}.$$

\section{Discussion}\label{sec-discuss}
A \emph{nearest neighbor tree}, introduced in \cite{KhanP06}, is any
tree that can be produced by Algorithm \ref{alg:nnt} shown below.
(Note that it always produces a tree.) Using the technique of
Section \ref{sec1}, we are able to get a simpler proof of the main
approximation result from \cite{KhanP06}.

\begin{algorithm}
\begin{algorithmic}
 \State each node $v$ is assigned a unique rank $r(v)$
 \For{each node $v$ such that $v$ is not the maximum-rank node
 {\bf (in parallel)}}
 \State
   let $w$ be a node such that
 $c(v, w) = \min\{c(v, v') \mid r(v') > r(v)\}$
 \State connect from $v$ to $w$ (i.e., add the edge $\{v, w\}$ to the
 tree)
 \EndFor
\end{algorithmic}
\caption{Algorithm for construction of a nearest neighbor
tree}\label{alg:nnt}
\end{algorithm}
\vspace{-0.15cm}

\begin{theorem}
The cost of any nearest neighbor tree $T$ is at most $O(\log n)$
times the cost of a minimum spanning tree. \end{theorem}
\begin{proof} Let $\lambda_j$ denote the number of edges in $T$ of
cost $j$ or more. Let $OPT$ be a minimum spanning tree and let $o$
be a depth-first search traversal of $OPT;$ it follows that $ \lVert
o \rVert \leq 2 c(OPT).$ As in the proof of Lemma \ref{lemma2}, for
any integer $j,$ we can partition $V$ into $\lVert o \rVert / j+1$
parts $P_i$ such that in each part, every pair of nodes is at most a
distance $j-1$ apart.

In each part $P_i,$ we claim that at most one node in $P_i$ tries to
form a connection of cost $j$ or greater. Indeed, only the
maximum-rank node $u$ in $P_i$ can do so, as all others can connect
to $u$ instead at cost at most $j-1.$ As before we find that
$\lambda_j \leq \lVert o \rVert / j$ which permits us to use the
same integral estimate as in the proof of Theorem \ref{thm1}. We get
\begin{equation}c(T) = \sum_j \lambda_j \leq (1 + \ln n) \lVert o \rVert =
O(\log n \cdot c(OPT)). \tag*{\qedhere}\end{equation}
\end{proof}

There is an interesting and difficult related problem which we were
unable to solve. Consider our original problem of counting the
number of steps taken by an agent executing the \NN\ heuristic ---
in other words, assume that $c$ is the distance function for some
(unweighted) graph. The costly \NN\ traversal of layered ring
graphs, and similarly the \NN\ traversal of the example from
\cite{hurkins}, both perform a lot of arbitrary tie-breaking. If we
break all ties randomly, then the performance seems to improve. Is
it possible that this would improve the approximation ratio of \NN\
to $O(1)$? (An observation in \cite{approxtsp} shows in the case of
edge-weighted graphs, random tie-breaking doesn't help.) Similarly,
when the edge-deleting adversary is not adaptive, does randomization
help in the distributed setting?

\newcommand{\noopsort}[1]{} \newcommand{\printfirst}[2]{#1}
  \newcommand{\singleletter}[1]{#1} \newcommand{\switchargs}[2]{#2#1}

\appendix
\section{Proof of Equation~\eqref{form}}\label{app:form}
A \emph{leg} of $L^i$ is an ordered pair $(a, b)$ such that $N^i(a,
b).$ The \emph{length} of that leg is $b-a.$

\begin{definition}
Let $S(k, t)$ denote the number of legs of length $2^t$ in $L^k.$
\end{definition}
The iterative construction of the graphs gives the following
recurrence relation.
\begin{enumerate}
\item $S(1, 0) = 2,$ and $S(1, 1)=S(1, 2)=\dotsb=S(1, m-1)=1.$
\item For $t>0$ and $k>1,$ we have $S(k, t)=\sum_{u>t} S(k-1, u).$
\item For $k>1,$ we have $S(k, 0)=S(k-1, 0)+2\sum_{u>0} S(k-1, u).$
\end{enumerate}

\begin{claim}
The solution of this recurrence relation for $S$ is
$$S(k, t) = \binom{m-t-1}{k-1} \textrm{ for } t>0; \quad S(k, 0) = 2\sum_{i=0}^{k-1} \binom{m-1}{i}.$$
\end{claim}
\begin{proof}
By using the identity
$$\sum_{z \leq A} \binom{z}{B} = \binom{A+1}{B+1},$$
it is easily verified that the claimed formulas satisfy conditions
(1)--(3).
\end{proof}

We have that $|L^k| = 1+\sum_{t \geq 0} S(k, t).$ We can simplify
the part of the sum with $t \geq 1$ since
$$\sum_{t \geq 1} S(k, t) = \sum_{t \geq 1} \binom{m-1-t}{k-1} =
\binom{m-1}{k}.$$ This observation leads to the following formula
for $|V(LR^k(m)|$ (note that we include the $2^m+1$ backbone nodes).
\begin{align*}
|V(LR^k(m)| &= 2^m + 1 + \sum_{j = 1}^k \left(1 + \sum_{t \geq 1}
S(j,
t) + S(j, 0)\right) \\
&= 2^m + k + 1 + \sum_{j=1}^k \left( \binom{m-1}{j} +
2\sum_{i=0}^{j-1} \binom{m-1}{i} \right) \\
&= 2^m + k + 1 + 2k\binom{m-1}{0} + \sum_{i=1}^{k}
(2k-2i+1)\binom{m-1}{i}.
\end{align*}

\section{Approximation Ratio Interpolation} \label{app-fooey}
We need to do some interpolation to show that
$$\lim_{n \to \infty} ar(n) / \log_2(n) = 1/2.$$
The problem is that as $m$ increases by one, the graphs
$LR^{(m-1)/(2+\delta)}(2^m)$ roughly double their number of
vertices, leaving a large gap. For this purpose, we may generalize
the construction of layered ring graphs in the following way. We
replace the size parameter $2^m$ by a size parameter $\nu,$ and no
longer insist that $L^i$-neighbors differ by a power of 2. We
redefine the layers in the following way.

\begin{definition}
Define the first layer $L^1$ as follows:
$$L^1 := \{0\} \cup \{\lceil{\nu/2^t}\rceil : t \geq 0\}.$$
For $i \geq 1,$ define the $(i+1)$st layer $L^{i+1}$ as follows:
$$L^{i+1} := \{0\} \cup \bigcup_{a, b : N^{i}(a, b)} \{a+\lceil(b-a)/2^t\rceil: t \geq 0\}.$$
\end{definition}

Having defined the layers, we define the layered ring graphs
$LR^k(\nu)$ using Definition \ref{lrdef} exactly as before. It is
straightforward to see that this indeed generalizes our previous
construction. That is, if $\nu = 2^m,$ then $LR^k(\nu) = LR^k(2^m)$
for all $k.$ We omit the straightforward proof of the following
claim.
\begin{claim} \label{daclaim} For fixed $k,$ $|V(LR^k(\nu))|+1 \leq |V(LR^k(\nu+1))| \leq
|V(LR^k(\nu))| + (k+1).$
\end{claim}

\begin{theorem}
The approximation ratio $ar(n)$ of \NN\ satisfies
\begin{equation}
\lim_{n \to \infty} ar(n) / \log_2(n) = 1/2.
\end{equation}
\end{theorem}
\begin{proof}
Considering Theorem \ref{thm:lowerbound}, we need to show that
$\liminf_{n \to \infty} ar(n) / \log_2(n) \geq 1/2.$ Let $\delta>0.$

For any $n,$ pick $m$ so that
\begin{equation}
\label{foog} |V(LR^{\lfloor(m-1)/(2+\delta)\rfloor}(2^m))| \leq n <
|V(LR^{\lfloor m/(2+\delta) \rfloor}(2^{m+1}))|.
\end{equation}

From Lemma \ref{lemma:lower} it follows that $m \sim \log_2 n.$ Fix
$k = \lfloor m/(2 + \delta) \rfloor.$ It also follows from Lemma
\ref{lemma:lower} that $LR^{k}(2^{m+1})$ has $o(n)$ non-backbone
vertices.

Pick the largest $\nu$ such that $|V(LR^k(\nu))| \leq n;$ by
Equation~\eqref{foog}, $\nu < 2^{m+1}.$ By Claim \ref{daclaim} it
follows that $LR^k(\nu)$ graph has $o(n)$ non-backbone vertices, and
hence that $\nu \sim n.$

From Claim \ref{daclaim} it also follows that by adding at most
$k+1$ vertices to $|V(LR^k(\nu))| \leq n$ we can obtain a graph on
exactly $n$ vertices, which we will call $G_n.$ Connect these new
vertices in a clique and connect them to $b_0$ and $\ell^1_\nu.$
There is an \NN\ traversal of $G_n$ where we visit the new clique
first, then the backbone, and then the layers in decreasing order;
the proof is analogous to Lemma \ref{mainclaim}. This \NN\ traversal
takes at least $\nu k \sim n \log_2 n / (2+\delta)$ steps, whereas a
hamiltonian circuit exists in $G_n,$ and so $ar(n) \gtrsim \log_2 n
/ (2+\delta).$ Taking $n \to \infty, \delta \to 0$ completes the
proof.
\end{proof}

\section{Distributed Restarting-\DFS\ is Slow}\label{app:dfs}
In this section we consider a version of \DFS\ that is adapted for
the distributed setting with edge failures, which is the network
model used in Section \ref{sec-appl}. Recall that edge failures are
allowed, but edge additions/restorations are forbidden. We consider
the following protocol for network traversal: the agent performs a
depth-first search, but whenever it is required to backtrack an edge
$e$ that has been deleted since the agent traversed $e$ in the
forwards direction, the agent begins a completely new depth-first
search. The algorithm terminates once a \DFS\ successfully completes
(i.e., returns to its originating vertex, and has explored all of
its neighbors). It is not hard to see that this algorithm will
eventually terminate successfully (i.e., the agent will have visited
all nodes of the connected component within which it lies).

There is a simple upper bound on the number of steps taken by this
protocol: the \DFS\ can restart at most $|E| = O(n^2)$ times since
there are at most $|E|$ edges that can be deleted, and each
individual \DFS\ takes at most $2(|V|-1) = O(n)$ steps, so the total
number of steps is at most $O(n^2 \cdot n) = O(n^3).$ We claim that
in fact $\Omega(n^3)$ steps can be taken in the worst case.

Here is the construction. Consider a graph that consists of two
cliques $C_1, C_2,$ each on $n/3$ nodes, joined by a path having
$n/3$ internal vertices. Fix a spanning tree $T$ such that no path
in $T$ contains $V(C_1)$ or $V(C_2)$; such a tree is easily seen to
exist for $n/3 > 3.$ Begin with the agent at some node $p \in C_1.$
Pick any edge $uv$ such that $\{u, v\} \subset C_2$ and $uv \not\in
T$; without loss of generality assume $v$ is not on the $p$-$u$ path
in $T.$ Have the agent walk along this path to $u,$ and then
traverse $uv.$ Then, delete $uv.$ By our choice of $T,$ the agent
will eventually need to backtrack to $p \in V(C_1);$ however, the
backtracking will first attempt to traverse $vu,$ causing a restart.
The agent is left at $v,$ and symmetrically to before, we pick any
remaining non-tree edge $u'v'$ in $C_1$ (with $v'$ not on the
$v$-$u'$ path in $T$), send the agent along $T$ to $u'$ and across
$u'v',$ and delete $u'v'.$ We repeat this process, sending the agent
between the cliques (and hence across the $n/3+1$-edge path) a total
of $|E \backslash T| = 2((n/3)^2-(n/3)+1)$ times, and thus using
$\Theta(n^3)$ steps.


\begin{thebibliography}{1}

\bibitem{alon92line}
N.~Alon and Y.~Azar.
\newblock On-line steiner trees in the euclidean plane.
\newblock In {\em Symposium on Computational Geometry}, pages 337--343, 1992.

\bibitem{ccps}
W.~J. Cook, W.~H. Cunningham, W.~R. Pulleyblank, and A.~Schrijver.
\newblock {\em Combinatorial Optimization}.
\newblock John Wiley \& Sons, Inc., New York, NY, USA, 1998.

\bibitem{hurkins}
C.~A.~J. Hurkens and G.~J. Woeginger.
\newblock On the nearest neighbor rule for the traveling salesman problem.
\newblock {\em Oper. Res. Lett.}, 32(1):1--4, 2004.

\bibitem{KhanP06}
M.~Khan and G.~Pandurangan.
\newblock A fast distributed approximation algorithm for minimum spanning
  trees.
\newblock In S.~Dolev, editor, {\em DISC}, volume 4167 of {\em Lecture Notes in
  Computer Science}, pages 355--369. Springer, 2006.

\bibitem{monnot}
J.~Monnot.
\newblock Approximation results toward nearest neighbor heuristic.
\newblock {\em Yugoslav Journal of Operations Research}, 12(1):11--16, 2002.

\bibitem{randalgs}
R.~Motwani and P.~Raghavan.
\newblock {\em Randomized Algorithms}.
\newblock Cambridge University Press, 2000.

\bibitem{approxtsp}
D.~J. Rosenkrantz, R.~E. Stearns, and P.~M. {Lewis II}.
\newblock An analysis of several heuristics for the traveling salesman problem.
\newblock {\em SIAM J. Comput.}, 6(3):563--581, 1977.

\end{thebibliography}
\end{document}